# Narrowband polarization-entangled photon pairs distributed over a WDM link for qubit networks

S. Sauge[1*], M. Swillo[1], S. Albert-Seifried[1], G. B. Xavier[1,2], J. Waldebäck[1], M. Tengner[1], D. Ljunggren[1], A. Karlsson[1]

[1]*Department of Microelectronics and Information Technology Royal Institute of Technology, KTH, Electrum 229, SE-16440 Kista, Sweden*
[2]*Center for telecommunication studies, Pontifical Catholic University, Rua Marquês de São Vicente, 225 Gávea, Rio de Janeiro, RJ, 22453-900, Brazil*
[*]sauge@kth.se

**Abstract:** We present a bright, narrowband, portable, quasi-phase-matched two-crystal source generating polarization-entangled photon pairs at 809 nm and 1555 nm at a maximum rate of $1.2 \times 10^6$ s$^{-1}$ THz$^{-1}$ mW$^{-1}$ after coupling to single-mode fiber. The quantum channel at 1555 nm and the synchronization signal gating the single photon detector are multiplexed in the same optical fiber of length 27 km by means of wavelength division multiplexers (WDM) having 100 GHz (0.8 nm) spacing between channels. This implementation makes quantum communication applications compatible with current high-speed optical networks.

## References and links

1. N. Gisin, G. Ribordy, W. Tittel, and H. Zbinden, "Quantum cryptography," Rev. Mod. Phys. **74**, 145-195 (2001).
2. A.K. Ekert, "Quantum cryptography based on Bell's theorem," Phys. Rev. Lett. **67**, 661 (1991).
3. I. Marcikic, H. de Riedmatten, W. Tittel, H. Zbinden, and N. Gisin, "Long-distance teleportation of qubits at telecommunication wavelengths," Nature **421**, 509-513 (2003).
4. H. Briegel, W. Dür, J. I. Cirac and P. Zoller, "Quantum repeaters: the role of imperfect local operations in quantum communication," Phys. Rev. Lett. **81**, 5932-5935 (1998).
5. P. G. Kwiat, K. Mattle, H. Weinfurter, A. Zeilinger, A. V. Sergienko, and Y. H. Shih, "New High-Intensity Source of Polarization-Entangled Photon Pairs," Phys. Rev. Lett. **75**, 4337-4340 (1995).
6. C. E. Kuklewicz, M. Fiorentino, G. Messin, F. N. C. Wong, and J. H. Shapiro, "High-flux source of polarization entangled photons from a periodically-poled KTiOPO$_4$ parametric downconverter," Phys. Rev. A **69**, 013807 (2004).
7. S. Tanzilli, H. De Riedmatten, W. Tittel, H. Zbinden, P. Baldi, M. De Micheli, D. B. Ostrowsky, and N. Gisin, "Highly efficient photon-pair source using a periodically poled lithium niobate waveguide," Electron. Lett. **37**, 26–28 (2001).
8. M. Pelton, P. Marsden, D. Ljunggren, M. Tengner, A. Karlsson, A. Fragemann, C. Canalias, and F. Laurell, "Bright, single-spatial-mode source of frequency non-degenerate, polarization-entangled photon pairs using periodically poled KTP," Opt. Express **12**, 3573-3580 (2004).
9. D. Ljunggren, M. Tengner, P. Marsden, and M. Pelton, "Theory and experiment of entanglement in a quasi-phase-matched two-crystal source," Phys. Rev. A **73**, 032326 (2006).
10. D. Ljunggren, and M. Tengner, "Optimal focusing for maximal collection of entangled narrow-band photon pairs into single-mode fibers," Phys. Rev. A **72**, 062301 (2005).
11. C. Liang, K. F. Lee, J. Chen, and P. Kumar, "Distribution of fiber-generated polarization entangled photon-pairs over 100 km of standard fiber in OC-192 WDM environment," postdeadline paper, Optical Fiber Communications Conference (OFC'2006), paper PDP35.
12. H. Hübel, M. R. Vanner, T. Lederer, B. Blauensteiner, A. Poppe, and A. Zeilinger, "High-fidelity transmission of polarization entangled qubits over 100 km of telecom fibers," (submitted to Opt. Express).
13. P.W. Shor, and J. Preskill, "Simple proof of security of the BB84 Quantum key distribution protocol", Phys. Rev. Lett. **85**, 441-444 (2000).
14. J. J. Xia, D. Z. Chen, G. Wellbrock, A. Zavriyev, A. C. Beal, and K. M. Lee, "In-band quantum key distribution (QKD) on fiber populated by high-speed classical data channels," Optical Fiber Communications Conference (OFC'2006), paper OTuJ7.

15. M. Fiorentino, G. Messin, C. E. Kuklewicz, F. N. C. Wong, and J. H. Shapiro, "Generation of ultrabright tunable polarization entanglement without spatial, spectral, or temporal constraints," Phys. Rev. A **69**, 041801 (2004).

## 1. Introduction

Entanglement is a fundamental resource in several quantum communication schemes used for quantum-key distribution (QKD) [1,2], teleportation [3] or quantum repeaters [4]. In the case of QKD, using pairs of entangled photons allows the exchange of random but yet correlated qubits used for the distribution of a cryptographic key between two distant parties (often referred to as Alice and Bob). In practical implementations, qubits can be encoded in the phase or polarization of single photons, which are entangled in that same degree of freedom.

Most sources of entangled photon pairs use spontaneous parametric down-conversion (SPDC). In the best known case [5], photon pairs are emitted along two cones with orthogonal polarizations. At the intersection of the cones, different polarizations cannot be distinguished and the two photons become entangled as a result. Moreover, by utilizing quasi-phase matched (QPM) periodically poled (PP) crystals [6] or waveguides [7], collinear emission can be achieved more easily to provide much greater overlap in the emission.

Besides from a high generation rate, a narrow bandwidth of the emitted pairs is a key requirement to increase transmission distance, as it will relax the need for compensating chromatic dispersion in the fiber, which otherwise will increase the uncertainty in the photon arrival time and thus the noise of the single photon detectors gated for longer times. Moreover, bridging the source atop current optical networks requires compatibility with Wavelength Division Multiplexing (WDM) environment operating typically with 100 GHz (0.8 nm) spacing between channels.

Following previous developments by our group in this field [8,9], we built a brighter and more narrowband source of polarization-entangled photon pairs based on collinear SPDC in two 50 mm long orthogonally oriented, periodically poled crystals. According to simulations performed on this source [10], the photon pair generation rate and the bandwidth expected with optimal focusing and single-mode emission are proportional to $L^{1/2}$ and $1/L$, respectively, where $L$ is the length of each crystal. Assuming optimal focusing, we can therefore only see advantages using long crystals. To our knowledge, this is the first time that such long crystals are used in this configuration.

While one motivation of this work was to test long crystals for the generation of narrowband photon-pairs, a second motivation was to distribute those narrowband pairs in a WDM environment, in order to test the compatibility of our source with current high-speed optical networks. At the same time, WDM provide the means to multiplex in the same optical fiber each qubit-encoded photon with the synchronization pulse needed to gate the single photon detector at the receiver side (Bob). Multiplexing would naturally limit the cost of deploying quantum cryptography, as no dedicated fiber would be needed for the quantum channel, which carries an intensity of only -100 dBm average power at 1 Mbps. However, about 100 dB isolation will be necessary to avoid cross-talk between the quantum and the synchronization channels. As will be shown, a channel spacing as low as 100 GHz (0.8 nm) between both channels can be achieved while keeping the required isolation. To the best of our knowledge, only one other group worldwide has also reported multiplexing of quantum and classical channels with such a narrow spacing [11].

This paper is organized as follows. In section 2, we present the new source used in a WDM environment. In section 3, we outline the results obtained when distributing entangled pairs over 27 km of standard SMF-28 single mode fiber and in section 4, we discuss future developments to increase transmission distance and key transmission rate.

## 2. The source

The portable source (45 cm × 60 cm) owned by Alice is presented in Fig. 1 (left side). The main element is the set of two orthogonally-oriented 50 mm-long bulk crystals made of MgO doped lithium niobate PPLN:MgO poled with a grating period of 7.5 µm (HC-Photonics). Crystals are pumped by a solid state diode laser operating at a wavelength of 532 nm (Cobolt). Once filtered from the residual light emitted at around 808 nm (BPF), the pump laser is polarized to 45º (HWP) and focused ($L_p$) such that each photon has an equal probability of down-converting either in the first or in the second crystal, respectively generating vertically (V) or horizontally (H) polarized photon pairs. The source has collinear emission at the non degenerate wavelengths of $\lambda_s$ = 809 nm for the signal (remaining at Alice) and $\lambda_i$ = 1555 nm for the idler (sent to Bob), allowing efficient single-photon detection and low attenuation in fibers at telecom wavelength, respectively (wavelength tuning can be achieved by varying the temperature of the brass ovens heating the crystals at around 100 ºC).

For the pump beam, we estimated a waist at the focus point of $2w_0$ = 150 µm, giving a Rayleigh range $z_0 = \pi w_0^2/\lambda$ of about 72 mm in PPLN:MgO, hence a $z_0/L$ ratio of 1.44. For the signal and idler beams, we measured a beam divergence $2\theta$ of ca. 9 mrad and 15 mrad respectively, leading to beam waists $w_0 = \lambda/\pi\theta$ of 110 µm at 809 nm and 130 µm at 1555 nm, which gives an average value of 120 µm.

After efficient blocking of the pump by a band-stop filter (BSF), signal and idler beams are spatially separated on a dichroic mirror (DM), collimated ($L_s$, $L_i$), filtered from residual pump light by means of long-pass filters (LPF) and eventually focused into single-mode fiber (SMF) at Alice's side. Due to the difference in beam divergences at 809 nm and 1555 nm after generation in the two crystals, we selected for each arm a collimating lens with a specific focal length ($f_s$ = 200 mm, $f_i$ = 150 mm) such that each collimated beam gets coupled with a focusing angle matching the numerical aperture of the fibers ($NA_s$ = 0.12, $NA_i$ = 0.13).

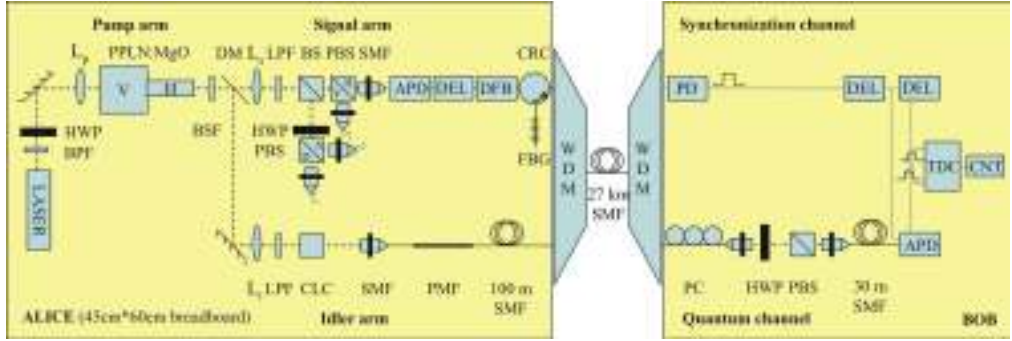

Fig. 1. Experimental setup. <u>ALICE</u>. *Pump arm*. LASER, 532 nm 50 mW pump laser; BPF, band-pass filter; HWP, half-wave plate; $L_p$, pump focusing lens ($f_p$ = 150 mm); BSF, 532 nm band-stop filter; DM, dichroic mirror. *Signal arm*. $L_s$, collimating lens for signal ($f_s$ = 200 mm); LPF, long-pass filter (Schott RG715); BS, 50/50 beam-splitter; PBS, polarizing beam-splitter; SMF, single-mode fiber coupler; APD, Si avalanche photo-detector; DEL, delay generator; DFB, distributed-feedback (trigger) laser at 1555.75 nm; CRC, circulator; FBG, Fiber Bragg grating. *Idler arm*. $L_i$, collimating lens for idler ($f_i$ = 150 mm); LPF, long-pass filter (Schott RG1000); CLC, calcite; PMF, polarization-maintaining fiber (3 meter-long). WDM, wavelength division multiplexer. <u>BOB</u>. *Synchronization channel*. PD, 1GHz InGaAs fiber optic photo-detector; TDC, time-discriminator-circuit; CNT, counter. *Quantum channel*. PC, fiber coupled polarization controller. APD, InGaAs avalanche photo-detector. Dashed lines, free-space; black lines, single-mode fiber; grey line, BNC cable. We use external (electronic) triggering (with a BNC cable connecting both single-photon detectors) when measuring the performances of the source without the 27 km fiber link.

Coupling to SMF provides spatial indistinguishability between H and V polarizations states and thus allows them to interfere in the diagonal basis, leading to entanglement. The resulting state is the Bell-state

$$|\Phi^\varphi\rangle = 1/\sqrt{2}\,(\,|V(\omega_s)\,V(\omega_i)\rangle + e^{i\varphi}\,|H(\omega_s)\,H(\omega_i)\rangle),\; \omega_i + \omega_s = \omega_p, \qquad (1)$$

where $\omega_p$, $\omega_s$ and $\omega_i$ are the pump, signal and idler frequencies respectively and $\varphi$ is the total phase difference of the two polarization components.

The polarization state of the 809 nm photon is measured locally at Alice's side while the 1555 nm photon is sent over optical telecom fiber to Bob, who performs his own polarization analysis. In the signal arm, the random choice of the two conjugate measurement bases (H/V for horizontal/vertical and D/A for diagonal/anti-diagonal) is made passively by a 50/50 beam-splitter (BS). For each basis, we use a Glan-Thomson prism (PBS) to separate both polarizations, with an extra half-wave plate (HWP) in the case of the D/A basis (to rotate D/A photons back to H/V before they enter the PBS). In the idler arm (after the fiber link), we use only one analyzer made of a PBS, and a motorized HWP for the purpose of recording two-photon visibility curves. We have also a polarization controller (PC) to align the polarization at the output of the fiber with respect to Bob's analyzer.

Chromatic dispersion leads to a group delay difference between signal and idler, for each polarization. After passing through (V) *and* (H) crystal, signal and idler (V) have a difference in group delays 11 ps larger than the one between signal and idler (H). Since the estimated coherence time of idler (V or H) is about 16 ps, the 11 ps delay between idler (V) and idler (H) leads to 75% lower visibility in the D/A basis than in the H/V basis. In order to restore indistinguishability, we slow down idler (V) with respect to idler (H) by transmitting them through a 25 mm long piece of birefringent calcite crystal (CLC). It introduces ca. 15 ps group delay difference between idler (H) and (V). We complemented CLC by a 3 meter-long piece of polarization maintaining fiber (PMF), which introduces a group delay difference between idler (H) and (V) of -4 ps (slow axis of PMF is aligned with fast axis of CLC). The reason to use PMF is to fine-tune the phase difference $\varphi$ of the two polarization components. This is achieved by applying a (local) mechanical strain on the fiber.

Temperature instabilities of the ovens heating the crystals can result in that $\varphi$ drifts over time through the change of refractive index. As $\varphi$ is also proportional to the crystal length, using long crystals imposes a more severe constraint on temperature stability to preserve the phase difference and therefore a high visibility in the diagonal basis. With 50 mm long crystals, a temperature drift of 0.1º C results in a relative phase shift between signal and idler polarizations of the order of $\pi$, leading, in the diagonal basis, to uncorrelated signal and idler polarizations. To enforce temperature stability to better than 0.1º C, it was sufficient to place both thermo-stabilized ovens inside an aluminum box protecting crystals from the airflow in the room.

On the transmitter side (Alice), we multiplex the quantum channel at 1555 nm with a synchronization channel at 1555.75 nm by means of a WDM add/drop module (Oplink). The synchronization signal is emitted by a DFB laser, which is triggered upon detection of an 809 nm photon. Each gating pulse (10 ns wide) is generated by switching off the fast in-built electro-absorption modulator keeping the laser at a minimum output power of about -10 dBm. The synchronization signal is shifted 50 ns in time behind the single-photon in order to minimize cross-talk, which is then caused mainly by fluorescence of the DFB laser at the wavelength of the single-photon. We filter this fluorescence from a level of -50 dBm down to below -100 dBm by means of a fiber Bragg grating (FBG) and two WDM filters (including the one used for multiplexing). This provides 20 and 40 dB attenuation, respectively. On the receiver side (Bob), we drop the synchronization signal with two cascaded WDM add/drop filters providing 40 dB isolation with the single-photon detector used at Bob's side. The synchronization pulses reflected off Bob's WDM are detected by a fast (~ 1 GHz) homebuilt photodiode (PD) operated right above its sensitivity threshold, around -23 dBm. Each optical pulse is converted into a TTL signal used to gate the single-photon detector at Bob's side.

The avalanche photo-detectors (APD) used are four Si-based APD (Perkin Elmer SPCM-AQR-14) for the signal side (809 nm), having a quantum efficiency $\eta_s = 60\%$, and a home-built InGaAs-APD (Epitaxx) module for the idler side (1555 nm), having $\eta_i = 18\%$ quantum efficiency. The latter is gated with 2.5 ns-long pulses, and to avoid after-pulsing effects it is used together with a hold-off circuit, which discards any trigger pulse arriving within 10 μs after the detector has been gated, corresponding to the mean lifetime of the trapped carriers in the semi-conductor photo-diode. In order to limit the noise of the InGaAs detector, we reduced its gate time artificially to discriminate between true counts (well localized in time) and dark ones (which occur with a higher probability as the time during which the detector is gated is increased). For that purpose, we built a time-discriminator circuit (TDC), which registers a count if the two TTL pulses output by the two APD upon (coincident) detection overlap. The TDC provides a gain of a few percents in the visibility when the overlap between the two TTL pulses is adjusted to be $\tau \approx 1.5$ ns (corresponding to the new effective gate time of the APD). When using TDC, about 20% of coincidence events also get discarded so we increase the pump power in order to reach the same coincidence rate as when TDC is not used, this time with a higher visibility. As a result, the single count rate at 809 nm is increased from $0.8 \times 10^6$ s$^{-1}$ to $1.1 \times 10^6$ s$^{-1}$ when TDC is used.

## 3. Experimental results

As shown in Fig. 2, the bandwidth $\Delta\lambda_i$ of the 1555 nm photons is 0.5 nm FWHM before coupling to the WDM filters (those have 0.2 nm flat-top bandwidth and 0.5 nm FWHM). Assuming a chromatic dispersion (*CD*) of 18 ps/nm/km for standard single-mode fiber, one could transmit 1555 nm photons over a distance $d \approx 150$ km while keeping a chromatic dispersion-induced time spread of the photon $\tau_{CD} = CD \times \Delta\lambda_i \times d$ below the gate time $\tau \approx 1.5$ ns of the InGaAs-APD. The signal at 809 nm has an expected bandwidth $\Delta\lambda_s$ below 0.15 nm, as derived from the conversion factor between signal and idler bandwidths $\Delta\lambda_i = (\lambda_i / \lambda_s)^2 \times \Delta\lambda_s \approx 3.7 \times \Delta\lambda_s$ (this formula can be obtained by writing that both signal and idler have the same coherence length, which is proportional to $\lambda^2/\Delta\lambda$).

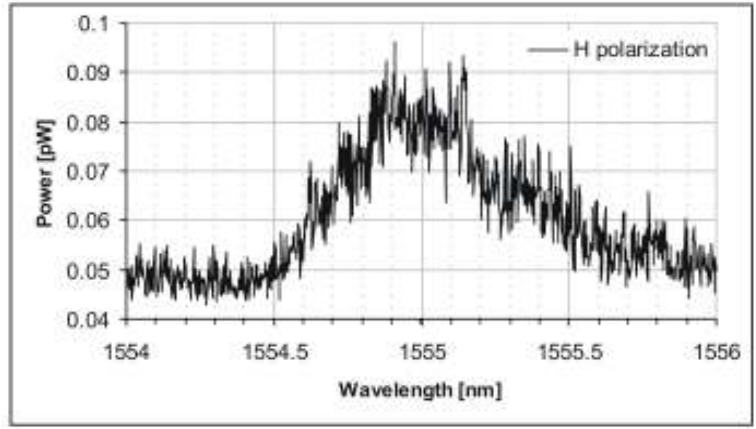

Fig. 2. 1555 nm spectrum for horizontal polarization without conditional gating at 809 nm. FWHM is about 0.5 nm. The background of 0.05 pW is coming from the noise of the detection system in the spectrograph (model Agilent 86140B).

The next three paragraphs present the performances of the source in the three following configurations: (i) we pump only one crystal and detect the 1555 nm photons directly after SMF coupling at Alice's side, (ii) we pump two crystals, use a fiber link of length 100 m and send the synchronization signal over a separate coaxial cable and (iii) we pump two crystals and multiplex both quantum and synchronization channels in the fiber link of length 27 km.

(i) We detected photon pairs at a maximum rate $R_c = 25 \times 10^3$ s$^{-1}$ when pumping one crystal only (V) at a power $P$ of around 3 mW. The single count rate at 809 nm was $R_s = 0.8 \times 10^6$ s$^{-1}$, yielding a conditional detection probability $R_c / R_s$ of about 3 % (TDC was not used in this particular case as it would lead to bias down our estimate of $R_c / R_s$). Taking into account the quantum efficiency $\eta_i = 18$ % of the InGaAs APD, we reach a corrected conditional detection probability of 16 %. The difference to an optimal value of around 90 % for single-crystal [10] is mostly due to losses arising from the requirement to get a balanced fiber in-coupling for both crystal sources in the D/A basis, as well as cumulated losses through optical components in the idler arm (mostly BSF and LPF, see Fig. 1). The rate of accidental counts, which we measured by random triggering of the InGaAs-APD, was $R_a = 0.9 \times 10^3$ s$^{-1}$, leading to a raw visibility $V_V = (R_c - R_a)/(R_c + R_a) = 93$ % at maximum coincidence count rate. Taking into account the quantum efficiency $\eta_s = 60$ % and $\eta_i = 18$ % of the two APD, we estimate that fibers carry photon pairs at a maximum rate $R_f \approx R_c /(\eta_s \eta_i) \times 1/P \times 1/(k\, \Delta\lambda_i) \approx 1.2 \times 10^6$ s$^{-1}$ THz$^{-1}$ mW$^{-1}$, where $k = 0.125$ THz nm$^{-1}$ is the conversion factor between nm and THz.

(ii) Figure 3 shows the polarization correlations between signal and idler photons with a fiber link of length 100 m. Visibility curves were obtained at a pump power of about 4 mW per crystal, single count rate of $1.1 \times 10^6$ s$^{-1}$ at 809 nm and only one WDM used with a bandwidth of 0.5 nm for 1555 nm photons. In comparison to (i), the coincidences rate $R_c$ dropped by a factor of about three, due to WDM coupling / filtering losses and insertion loss into the free-space analyzer used at Bob's side. Raw visibilities for each of the four polarization states (H/V and D/A) of the signal were $V_H = 94$ %, $V_V = 90$ %, $V_D = 87$ % and $V_A = 89$ %, respectively. The drop in the visibility (from 93 % to 90 % for $V_V$) despite the use of TDC is mainly due to non-ideal polarization alignment. Random fluctuations of the polarization after 100 m of fiber also affect how stable the visibility remains over time. The quantum bit error rate (QBER), given by the averaged ratio $R_a / R_c$ over all four bases, amounts to about 5 %.

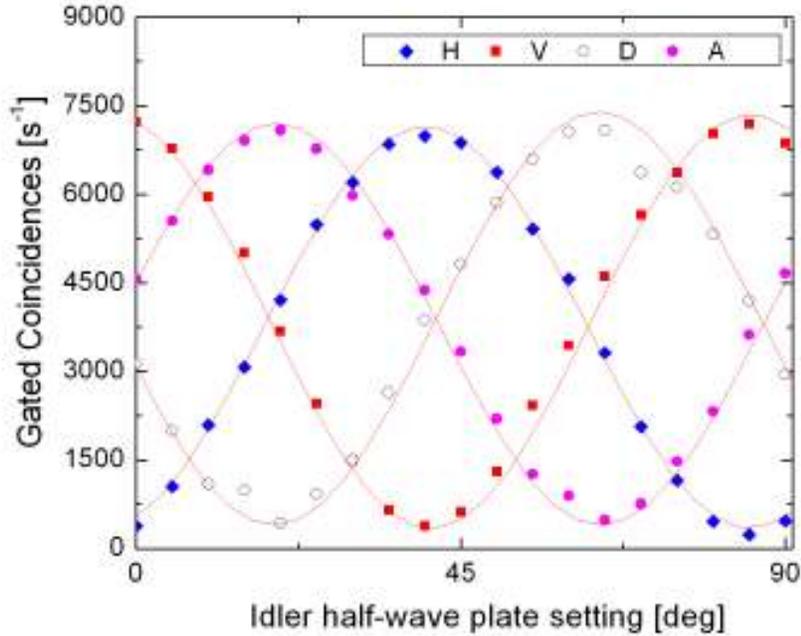

Fig. 3. Coincidence rate after 100 m of fiber as a function of idler polarization for each of the 4 polarization states (H/V and D/A) of the signal.

(iii) Figure 4 shows visibility curves when entanglement is distributed over 27 km of SMF fiber (FiberCore). Curves were obtained at a pump power of about 4 mW per crystal, single-count rate of $1.1 \times 10^6$ s$^{-1}$ at 809 nm and all four WDM filters inserted. The coincidences rate dropped to $1.1 \times 10^3$ s$^{-1}$, mainly due to the attenuation by the 27 km fiber link added (-6 dB) and insertion losses into the 3 other WDM (-3 dB). Raw visibilities decreased down to $V_H$ = 85 %, $V_V$ = 85 %, $V_D$ = 83 % and $V_A$ = 85 %, respectively, mainly due to a residual leak of the trigger laser into the quantum channel, which could be reduced by using larger channel spacing. The moderate drop in the visibility also suggests that polarization fluctuations do not increase significantly with longer fiber links. This is supported by [12] up to 100 km of fiber at least. From our observations, a quantum key distribution experiment could be run over a few minutes without need for active polarization control. The QBER for this measurement was about 8 %, still below the 11 % limit allowing secure QKD [13].

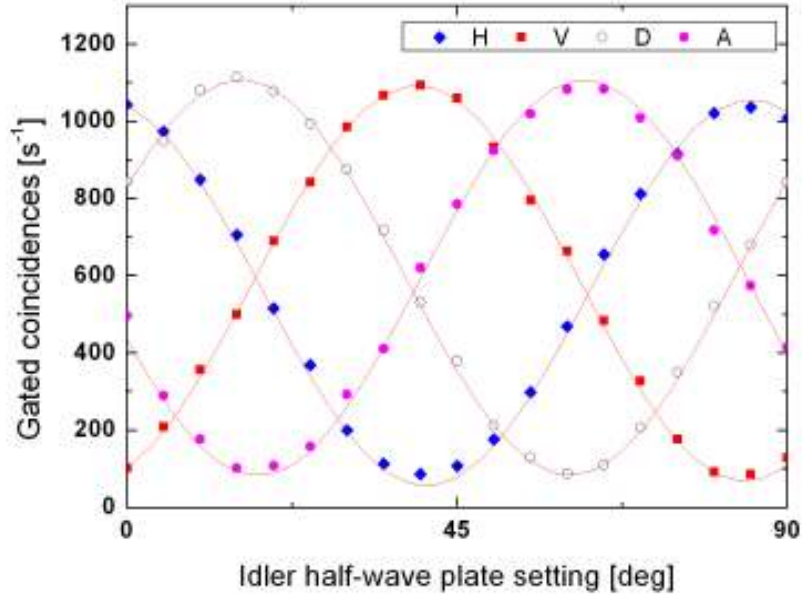

Fig. 4. Coincidence rate after 27 km of fiber as a function of idler polarization for each of the 4 polarization states (H/V and D/A) of the signal.

## 4. Conclusions

In this paper, we implemented the use of long periodically-poled crystals to increase the brightness and narrow the bandwidth of entangled photon pairs generated by a portable source at 809 nm and 1555 nm. With a bandwidth of only 0.5 nm at 1555 nm, photons can be transmitted over 150 km of single-mode fiber without the need to compensate for chromatic dispersion. Such a narrow bandwidth also allows multiplexing of the quantum and synchronization channels in a WDM environment with the attractive prospect of bridging the source atop current optical networks without the need for a dedicated fiber carrying the single photons. With only 0.8 nm separation between quantum and synchronization channels, we detected up to $1.1 \times 10^3$ s$^{-1}$ photon pairs with a raw visibility of 85 % in all measurement bases after transmission over 27 km of single mode fiber. With the current isolation of about 100 dB achieved between adjacent channels, we expect that the quantum channel could also be multiplexed with a classical data channel [14]. Investigation is under way.

Long crystals, however, will put strong constraints on the temperature stability of the crystals, as temperature variations will affect the refractive indexes, thus the output state phase factor and the quality of entanglement. From this point of view, a one-crystal source pumped both ways from opposite sides would make the set-up completely insensitive to temperature drifts of the heated crystal [15]. However, long term stability of the source will still require active control of the polarization, which should need to be monitored every few minutes.

**Acknowledgments**

We would like to thank Pr. Gunnar Björk for his careful reading of the manuscript; our colleagues from Acreo, Drs. Pierre-Yves Fonjallaz and Anders Djupsjöbacka, who provided the DFB laser at 1555.75 nm, the set of four WDM and the 1 GHz photodiode; our colleagues Drs. Andreas Poppe and Hannes Hübel from the Faculty of Physics in Vienna for the 27 km fiber spool and for numerous insightful discussions regarding our two sources. We also acknowledge support from HC Photonics for the periodically poled crystals made of LN:MgO as well as support from Cobolt AB for the lending of a replacement single-mode laser at 532 nm. This work is supported by the European Commission through the integrated project SECOQC (Contract No IST-2003-506812) and by the Swedish Foundation for Strategic Research (SSF). Additionally, G. B. Xavier acknowledges the financial support provided by the Brazilian agencies Capes and CNPq.